\documentclass[10pt]{article}
\usepackage{amsmath,amsthm,amssymb,amscd}
\usepackage{xurl}
\usepackage{hyperref}
\usepackage{geometry}
\usepackage{tikz-cd}
\usepackage{booktabs}
\usepackage{longtable}
\usepackage{graphicx}
\usepackage{float}
\usepackage[section]{placeins}
\usepackage{needspace}


\usepackage{amsmath}
\usepackage{amssymb}
\usepackage{booktabs}
\usepackage{xcolor}

\definecolor{Maroon}{RGB}{128,0,0}

\usepackage{multirow}
\usepackage{tikz}
\usetikzlibrary{calc,positioning,3d}
\usepackage[normalem]{ulem}
\usepackage{imakeidx}
\makeindex
\usepackage{subcaption}
\usepackage{caption}
\usepackage{makecell}

\theoremstyle{definition}

\title{Capturing Coordination Core Topology in Metal--Organic Frameworks Using Persistent Mayer Homology}

\author{
Caleb Simiyu Khaemba$^{1}$,
Chun-Long Chen$^{2}$%
\thanks{Corresponding author: chunlong.chen@pnnl.gov},
and Guo-Wei Wei$^{1,3}$%
\thanks{Corresponding author: guowei.wei@uga.edu}
\\[5pt]
\small $^{1}$Department of Mathematics, University of Georgia,
Athens, GA 30602, USA.
\\
\small $^{2}$Physical Sciences Division, Pacific Northwest National Laboratory,
Richland, Washington 99354, USA.
\\
\small $^{3}$Department of Biochemistry and Molecular Biology,
University of Georgia, Athens, GA 30602, USA.
}

\begin{document}

\maketitle

\begin{abstract}
Local coordination cores are critically important in metal–organic frameworks (MOFs) and polyoxometalates (POMs), as they govern the fundamental chemical and structural properties of these materials. Existing methods, including element-resolved persistent homology (ERPH), have limited ability to translate local coordination cores into quantitative, scalable, and interpretable information. Here, we introduce persistent Mayer homology (PMH) to enable the detection of the geometric variability invisible by classical topology. By focusing on coordination cores formed by periodic metal-centered heavy-atom, carbon, and donor channels, PMH distinguishes different arrangements of the local coordination environment, linker framework, and potential coordinating atoms. We examine explainable PMH of orders $N=3$ and $N=5$ and separately combine each with ERPH, which characterizes the multiscale organization of chemically defined element groups across the extended framework, to predict eight nitrogen- and oxygen-related adsorption and self-diffusion properties. The resulting coordination-core PMH representations consistently improve predictive performance over ERPH alone, with the third-order Mayer representation providing the strongest overall performance. These results demonstrate that the inner coordination configurations captured by PMH contain essential structural information that is not fully represented by conventional element-resolved topological descriptors. This finding provides a quantitative interpretable topological framework for understanding local coordination environments and may facilitate the rational design of MOFs and POM-based materials.
\end{abstract}


%\section{Introduction}
%\label{sec:introduction}

Connecting atomic structure to macroscopic function is a central objective of materials science. Among crystalline porous solids, metal--organic frameworks (MOFs) provide an exceptionally programmable materials platform because metal ions or clusters and organic linkers can be assembled into periodic structures with widely varying composition, connectivity, and pore geometry \cite{aunan2021modulation,bennett2021changing,moosavi2020}. This control has made MOFs important candidates for gas storage, molecular separation, and catalysis \cite{siegelman2021porous,altintas2021machine,chong2020}. Polyoxometalate (POM) based MOFs provide a particularly important example of the multiscale structural complexity addressed in this work, where inorganic clusters can be incorporated into extended porous frameworks \cite{chen2005co2}. POMs are discrete stoichiometric metal oxide clusters with rich electronic properties and the ability to undergo reversible redox transformations while maintaining their structural integrity \cite{papaconstantinou1967heteropoly,katsoulis1998survey,long2010polyoxometalates,song2012recent}. These characteristics have motivated their use in microelectronic and neuromorphic systems, including resistive switching and reservoir computing applications \cite{yang2023old,zhang2022,moors2024capacitor,akai2022performance}. Incorporating POMs into porous frameworks combines these distinctive electronic and chemical properties with the high surface area, tunable porosity, and structural periodicity of MOFs \cite{li2019recent,du2014recent,zhang2021pommof}. POM based MOFs therefore present a particularly rich multiscale structural problem: their behavior can depend on the internal organization and coordination of the multimetal oxide core as well as its connection to the surrounding periodic framework and pore environment. Capturing both scales is important because structural differences within the inorganic core may not be adequately represented by global pore information alone, while a description restricted to the core cannot capture the extended environment through which molecules adsorb and move. The same chemical and structural freedom, however, creates a materials space too large to examine systematically by synthesis and measurement alone \cite{moosavi2020,jablonka2020}. High throughput molecular simulation has therefore become an important component of MOF discovery, enabling large structural collections to be screened before experimental testing \cite{boyd2019,daglar2020}. Yet repeated adsorption and transport calculations remain costly, particularly when diffusion requires long molecular dynamics trajectories \cite{sharp2021nanoconfinement,lim2022}. Data driven surrogate models are consequently needed to relate crystal structure to adsorption affinity, uptake, and molecular mobility at substantially lower computational cost \cite{altintas2021machine,daglar2022membranes}.

The main difficulty is not only obtaining property labels, but also representing the structural information that governs them. Adsorption is influenced by the chemical identity, charge distribution, and geometry of binding sites, whereas capacity and diffusion also depend on accessible volume, limiting apertures, pore shape, and the connectivity of periodic channels \cite{ji2020porechemistry,anderson2018,sharp2021nanoconfinement}. These contributions are coupled: their relative importance changes with adsorbate identity, temperature, pressure, and loading, while confinement modifies both host--guest and guest--guest interactions \cite{daglar2020,siegelman2021porous}. Polyoxometalate based MOFs make this representation problem especially clear because a structurally complex metal oxide core can contain multiple metal centers and chemically distinct coordination environments within the same extended porous material \cite{du2014recent,zhang2021pommof,ebrahimi2024pommofs}. A representation based only on bulk pore statistics may therefore overlook configurations that distinguish the inner coordination cores, whereas a description restricted to nearby atoms may miss the extended channels through which molecules enter and move. The available data add a second difficulty. Adsorption and diffusion labels require grand canonical Monte Carlo or molecular dynamics calculations and can depend on structural preparation, force fields, charge assignment, and simulation protocol \cite{orhan2022,daglar2020,lim2022}. Experimental and computed MOF databases are also uneven in their chemical coverage, which can bias learned structure--property relationships and reduce transferability \cite{chung2019core,moosavi2020,nandy2022stable}. An effective representation must therefore preserve chemical identity, coordination structure, and periodic pore organization while remaining useful with moderate and unevenly distributed labeled data.

Previous machine learning studies have addressed parts of this problem through complementary representations. Geometric, chemical, and energy descriptors provide compact summaries of pore diameter, surface area, accessible volume, density, elemental composition, and guest interaction energy, and they have enabled rapid prediction of adsorption, selectivity, and diffusivity \cite{jablonka2020,altintas2021machine,orhan2022,yan2022}. These variables are interpretable and inexpensive, but a fixed set of summaries can suppress detailed coordination geometry and the arrangement of connected pores. Crystal graph models reduce dependence on manually chosen descriptors by learning chemical and geometric interactions from atomic neighborhoods, and have been applied to electronic and stability properties of large MOF collections \cite{rosen2021,nandy2022stable}. Transformer models enlarge the accessible structural context and use pretraining to transfer information from extensive porous material collections to tasks with fewer labels \cite{kang2023,park2023,wang2024}. Their predictive power is valuable, although the learned variables are not always readily associated with definite atomic configurations. Mathematical representations provide a complementary route in which geometry is converted into multiscale mathematical invariants \cite{nguyen2020review}. For example, persistent homology records the formation and disappearance of connected components, loops, and cavities \cite{carlsson2009topology}. It has been used to characterize the pore geometry in nano materials \cite{xia2015persistent} and predict adsorption in nanoporous materials \cite{lee2017,townsend2020molecular,krishnapriyan2021}. Element-specific persistent homology \cite{cang2018integration} extends this description by retaining chemical distinctions that ordinary geometric summaries discard \cite{jiang2021topological,chen2025cstl}. Interaction topology \cite{liu2025persistent} and commutative algebra \cite{suwayyid2025persistent} representations further organize chemical groups and their relations, connecting periodic pore organization with coordination motifs \cite{chen2025itt,khaemba2026}. These developments show the value of chemically informed mathematical descriptors, and recent analysis has emphasized the broader importance of topology in MOF characterization \cite{glasby2024topological,wee2025review}. Nevertheless, ordinary persistent homology is governed by the single boundary relation $d^{2}=0$. It does not explicitly examine the additional boundary relations available within the higher order simplicial organization of metal coordination cores.

In this study, interpretable persistent Mayer homology (PMH) \cite{shen2024} is used to capture the inner coordinating configurations of MOF metal cores. PMH is a generalization of the classical persistent homology by allowing high-order differentiations $d^{N}=0,~N\geq2$. When $N>2$, PMH is sensitive to the geometric changes that are not detectable by persistent homology \cite{su2025topological}. In our approach, the periodically expanded MOF is first separated into chemically defined element groups, and persistent homology is computed on their $\alpha$-complex filtrations to describe the multiscale organization of the pore structure. Each metal center is then represented by three periodic coordination channels within $10~\text{\AA}$: all heavy atoms, carbon atoms, and donor atoms. Persistent Mayer homology replaces $d^{2}=0$ with $d^{N}=0$ and yields the indexed groups $H_{k,q}$ for $q=1,\ldots,N-1$ \cite{shen2024}. Their Mayer Betti curves give several algebraic views of how vertices, edges, triangles, and higher dimensional simplices are arranged within each coordination core \cite{feng2025mayer}. We examine Mayer orders $N=3$ and $N=5$ and combine each coordination core representation with the element resolved persistent homology (ERPH) representation. The resulting representations connect chemically resolved pore organization with the internal simplicial configurations of the metal cores. They are evaluated on eight oxygen and nitrogen adsorption and diffusion properties derived from experimentally reported MOF structures \cite{chung2019core,orhan2022,chen2025cstl}.

The remainder of this paper is organized as follows. The Results and Discussion section presents the ERPH and coordination-core PMH representations, evaluates their performance on the eight adsorption and self-diffusion tasks, and discusses the additional topological invariants provided by persistent Mayer homology and their chemical interpretation. The Methods section describes the datasets, mathematical constructions, feature generation, and predictive modeling. Additional ablation results, implementation details, and mathematical properties of the coordination-core representation are provided in the Supporting Information.

\section*{Results and Discussion}
\label{sec:results}

\subsection*{Topological representation of MOFs}
\label{subsec:topological_representation}

\begin{figure}[!t]
\centering
\includegraphics[width=\linewidth]{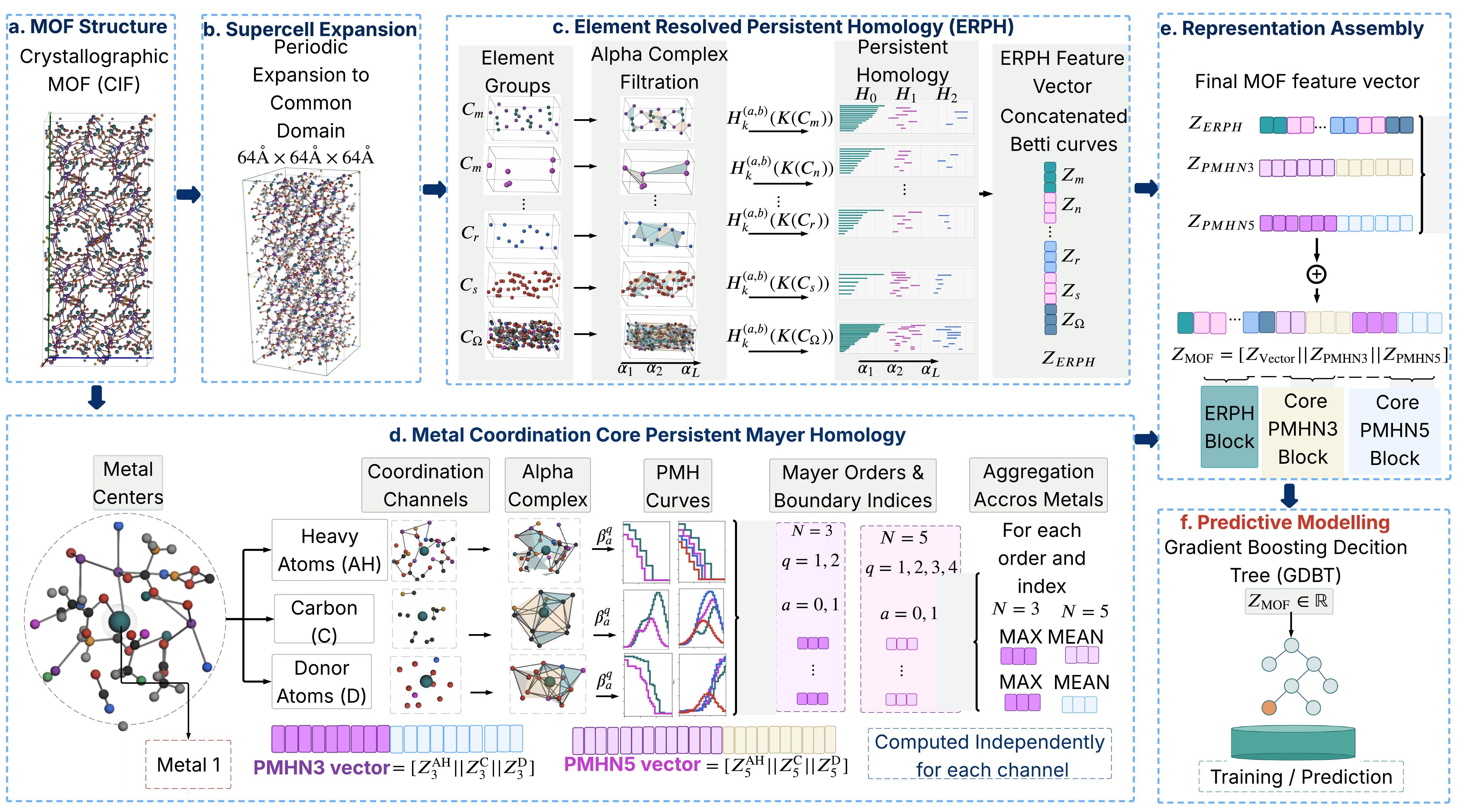}
\caption{MOF representation and prediction workflow. \textbf{a.} Crystallographic MOF structure obtained from a crystallographic information file. \textbf{b.} Finite periodic expansion with lattice-vector lengths approaching $64~\text{\AA}$ while preserving the original lattice geometry. \textbf{c.} ERPH computed from the element groups $C_m$, $C_n$, $C_o$, $C_p$, $C_q$, $C_r$, $C_s$, and $C_{\Omega}$. \textbf{d.} Persistent Mayer homology computed independently for the all-heavy-atom, carbon, and potential-donor coordination channels. Mayer Betti curves for $N=3$ and $N=5$ are summarized across distinct coordination-core classes using pointwise maximum and arithmetic mean. \textbf{e.} Assembly of the ERPH, PMHN3, and PMHN5 feature blocks. The principal models combine ERPH separately with each Mayer block; full concatenation is retained as an internal ablation. \textbf{f.} Gradient-boosted decision tree training and property prediction.}
\label{fig:mof_feature_workflow}
\end{figure}

Figure~\ref{fig:mof_feature_workflow} summarizes the complete representation and prediction workflow. Starting from a crystallographic MOF structure, the method constructs two complementary mathematical descriptions. Element-resolved persistent homology (ERPH) captures the chemical organization and topology of the periodically expanded structure, whereas metal-coordination-core persistent Mayer homology captures the internal configurations of atoms surrounding the metal centers.

Each MOF was read from its crystallographic information file and periodically expanded to approximately $64~\text{\AA}$ along each lattice direction (Fig.~\ref{fig:mof_feature_workflow}a,b). This provides comparable spatial coverage for ERPH while preserving the original lattice geometry and atomic coordinates; the structures are neither rescaled nor embedded in identical cubic cells. ERPH was evaluated over $\alpha=0,\dots,24.9~\text{\AA}$. The expansion supplies periodic atomic repetitions across this filtration range and reduces, but does not eliminate, finite-boundary effects \cite{khaemba2026}. The supercell atoms are separated into the chemically defined groups listed in Table~\ref{tab:element_categories}. The two metal groups are evaluated independently but are shown collectively as $C_m$ in Fig.~\ref{fig:mof_feature_workflow}c. The set $C_{\Omega}$ contains all atoms and provides an unpartitioned structural description. A separate $\alpha$-complex filtration is constructed for each point cloud, and persistent homology is computed in dimensions $H_0$, $H_1$, and $H_2$. Sampling the resulting Betti curves produces 6,750 ERPH features. The second representation is centered on the metal coordination cores (Fig.~\ref{fig:mof_feature_workflow}d). A metal anchor is any atom belonging to either metal group $C_m^{(1)}$ or $C_m^{(2)}$. For each anchor, periodic neighbors within $10~\text{\AA}$ are assigned to the all-heavy-atom, carbon, and potential-donor channels. A coordination core is therefore the periodic point cloud within $10~\text{\AA}$ of a metal anchor, not a bond-defined first coordination shell. The anchor is retained in every channel. Point clouds with the same anchor identity and the same canonical, element-labeled pairwise-distance code are assigned to one coordination-core class. Persistent Mayer homology is computed once for each distinct class at Mayer orders $N=3$ and $N=5$. Pointwise maximum and unweighted arithmetic mean summarize the distinct classes in each channel; repeated equivalent copies do not receive additional weight. The PMHN3 and PMHN5 blocks contain 960 and 1,920 features, respectively. The principal models combine ERPH separately with PMHN3 or PMHN5. Full feature concatenation and equal-weight prediction consensus are retained as internal ablations (Fig.~\ref{fig:mof_feature_workflow}e). A gradient-boosted decision tree then predicts the eight MOF properties (Fig.~\ref{fig:mof_feature_workflow}f).

\begin{table}[!t]
\centering
\caption{Element groups used to construct the ERPH representation. The two metal groups are evaluated independently and are shown collectively as $C_m$ in Fig.~\ref{fig:mof_feature_workflow}.}
\label{tab:element_categories}
\begin{tabular}{p{0.30\linewidth} p{0.49\linewidth} c}
\toprule
\textbf{Element group} & \textbf{Included elements} & \textbf{Notation} \\
\midrule
Main-group metals
& Alkali metals, alkaline-earth metals, and other main-group metals
& $C_m^{(1)}$ \\
Transition and inner-transition metals
& Transition metals, lanthanoids, and actinoids
& $C_m^{(2)}$ \\
Metalloids
& B, Si, Ge, As, Sb, Te, Po, and At
& $C_n$ \\
Halogens
& F, Cl, Br, and I
& $C_o$ \\
Hydrogen
& H
& $C_p$ \\
Carbon
& C
& $C_q$ \\
Nitrogen and phosphorus
& N and P
& $C_r$ \\
Oxygen, sulfur, and selenium
& O, S, and Se
& $C_s$ \\
Complete structure
& All atoms
& $C_{\Omega}$ \\
\bottomrule
\end{tabular}
\end{table}

\begin{figure}[!t]
\centering
\includegraphics[width=0.95\linewidth]{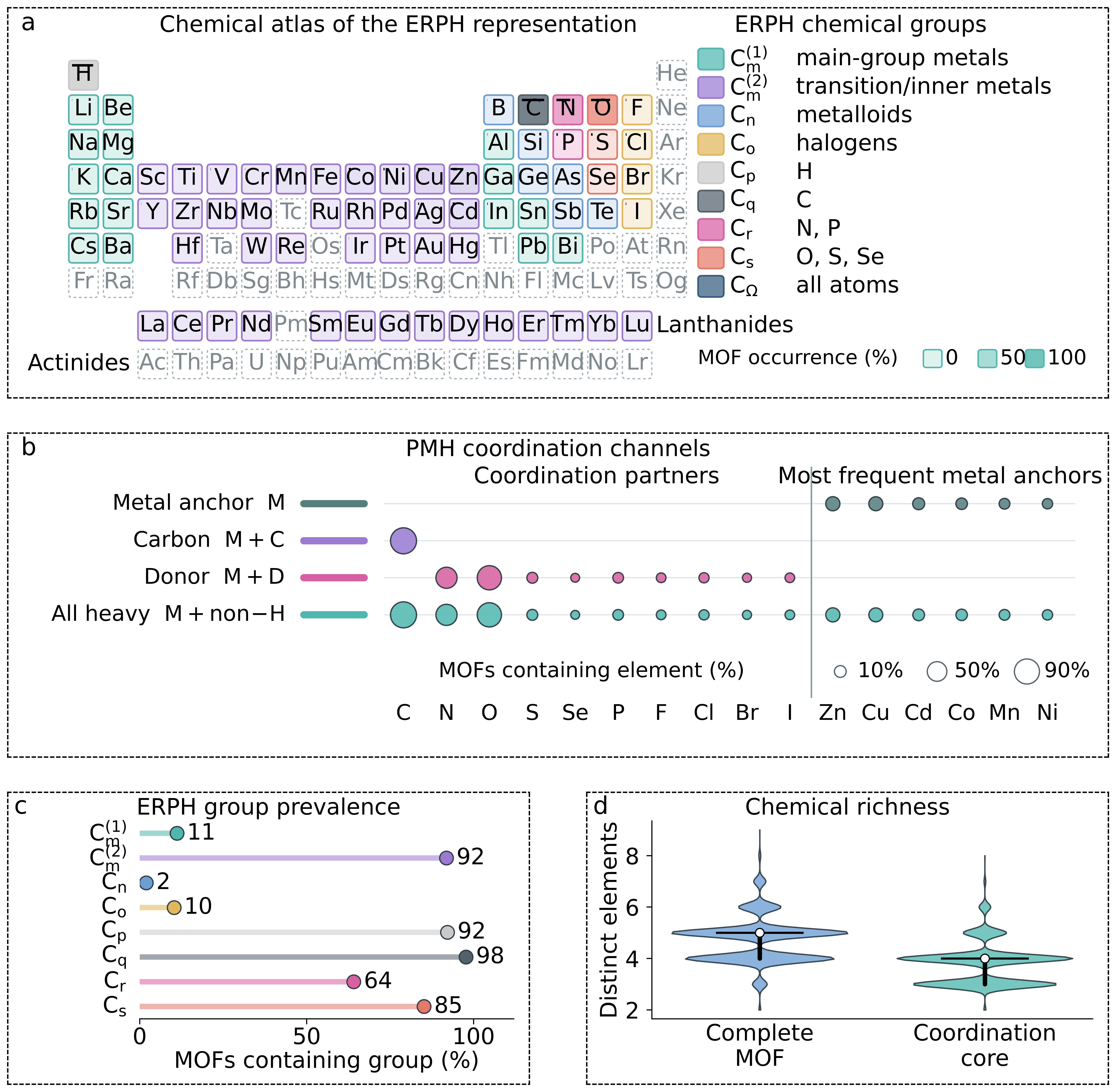}
\caption{Chemical coverage of the benchmark MOFs and its relationship to the ERPH and metal-coordination-core representations. \textbf{a.} Periodic-table map of element occurrence across the MOFs used in the eight prediction tasks. Each element is assigned to one of the ERPH groups $C_m^{(1)}$, $C_m^{(2)}$, $C_n$, $C_o$, $C_p$, $C_q$, $C_r$, or $C_s$, while $C_{\Omega}$ represents the complete structure containing all atoms. The color identifies the corresponding ERPH group, and the color intensity indicates the percentage of MOFs containing that element. Faded boxes with dashed borders indicate elements not observed in the benchmark structures. \textbf{b.} Chemical occurrence within the metal-coordination-core construction. The rows show metal-anchor occurrence followed by the carbon, potential-donor, and all-heavy-atom feature channels. The metal anchor is included in each of the three feature channels but does not form a separate channel. Bubble area is proportional to the percentage of MOFs containing the corresponding element in that channel. \textbf{c.} Percentage of benchmark MOFs containing at least one element from each ERPH group. \textbf{d.} Distribution of the number of distinct elements in the complete MOF structures and in their metal coordination cores. All percentages are calculated over the union of unique MOF identifiers appearing in the eight property datasets.}
\label{fig:chemical_atlas}
\end{figure}

Figure~\ref{fig:chemical_atlas} provides a chemical summary of the benchmark MOFs and shows how their elemental composition is incorporated into the two complementary mathematical representations. ERPH describes chemical and topological organization across the periodically expanded structure, whereas PMH describes the internal geometry of metal coordination cores through the carbon, potential-donor, and all-heavy-atom channels. The chemical distribution in Fig.~\ref{fig:chemical_atlas} shows that the benchmark set contains broad but uneven elemental coverage. Transition and inner-transition metals, $C_m^{(2)}$, occur in approximately $92\%$ of the MOFs, whereas main-group metals, $C_m^{(1)}$, occur in approximately $11\%$. Carbon is present in approximately $98\%$ of the structures, hydrogen in $92\%$, oxygen, sulfur, or selenium in $85\%$, and nitrogen or phosphorus in $64\%$. Halogens and metalloids are less frequent, occurring in approximately $10\%$ and $2\%$ of the MOFs, respectively. These differences show that the ERPH representation covers both the dominant chemical constituents and the less common element groups that may distinguish smaller subsets of structures. The coordination-channel analysis provides a complementary description of the metal-centered atomic configurations. Carbon, nitrogen, and oxygen are the most common coordination partners, while Zn, Cu, Cd, Co, Mn, and Ni are the most frequent metal anchors. The complete MOF structures generally contain more distinct elements than their corresponding coordination cores because the cores retain only the atoms within the local neighborhood of each metal center. Thus, ERPH describes chemical organization across the periodically expanded structure, whereas PMH concentrates on the internal configurations of the metal coordination cores. Their combination retains information from these two structural descriptions without treating them as identical.

\subsection*{Adsorption and self-diffusion prediction}
\label{subsec:adsorption_results}
\label{subsec:diffusion_results}

Table~\ref{tab:property_benchmark} compares the two proposed representations with the descriptor-based, MOFTransformer, and PMTransformer benchmark results \cite{orhan2022,kang2023,park2023}. ERPH combined with PMH at $N=3$ gives the highest reported mean $R^{2}$ for all four adsorption properties. The mean $R^{2}$ values are $0.84$ and $0.88$ for the nitrogen and oxygen Henry constants and $0.83$ and $0.90$ for nitrogen and oxygen uptake, respectively. The corresponding mean uptake MAEs are $4.32\times10^{-2}$ and $3.83\times10^{-2}$~mol/kg, both below the available transformer benchmark values. Because the external studies used different sample counts and data partitions, their results serve as contextual benchmarks rather than direct paired comparisons. The direct comparison among the representations developed here is reported in Table~\ref{tab:internal_ablation}.

\begin{table}[!t]
\centering
\caption{Comparison of the proposed representations with benchmark results for nitrogen and oxygen adsorption and self-diffusion in MOFs.}
\label{tab:property_benchmark}
\label{tab:adsorption_benchmark}
\label{tab:diffusion_benchmark}
\setlength{\tabcolsep}{3.5pt}
\renewcommand{\arraystretch}{1.04}
\small

\begin{tabular*}{\linewidth}{
@{\extracolsep{\fill}}
l l c c c c c
@{}
}
\toprule
Property
& Metric
& \shortstack{ERPH $+$ PMH\\$N=3$}
& \shortstack{ERPH $+$ PMH\\$N=5$}
& \shortstack{Descriptor based\\\cite{orhan2022}}
& \shortstack{MOFTransformer\\\cite{kang2023}}
& \shortstack{PMTransformer\\\cite{park2023}} \\
\midrule

\multicolumn{7}{l}{\textbf{Adsorption properties}} \\
\midrule

\multirow{3}{*}{Henry constant, $\mathrm{N_2}$}
& $R^{2}$ & $\mathbf{0.84}$ & $0.83$ & $0.70$ & -- & -- \\
& MAE & $\mathbf{4.41\times10^{-7}}$ & $4.56\times10^{-7}$ & -- & -- & -- \\
& RMSE & $\mathbf{6.64\times10^{-7}}$ & $6.82\times10^{-7}$ & $8.94\times10^{-7}$ & -- & -- \\
\addlinespace[2pt]

\multirow{3}{*}{Henry constant, $\mathrm{O_2}$}
& $R^{2}$ & $\mathbf{0.88}$ & $0.86$ & $0.74$ & -- & -- \\
& MAE & $\mathbf{4.33\times10^{-7}}$ & $4.56\times10^{-7}$ & -- & -- & -- \\
& RMSE & $\mathbf{6.61\times10^{-7}}$ & $6.99\times10^{-7}$ & $9.60\times10^{-7}$ & -- & -- \\
\addlinespace[2pt]

\multirow{3}{*}{$\mathrm{N_2}$ uptake}
& $R^{2}$ & $\mathbf{0.83}$ & $0.82$ & $0.71$ & $0.78$ & -- \\
& MAE & $\mathbf{4.32\times10^{-2}}$ & $4.49\times10^{-2}$ & -- & $7.10\times10^{-2}$ & $6.90\times10^{-2}$ \\
& RMSE & $\mathbf{6.63\times10^{-2}}$ & $6.82\times10^{-2}$ & $8.62\times10^{-2}$ & -- & -- \\
\addlinespace[2pt]

\multirow{3}{*}{$\mathrm{O_2}$ uptake}
& $R^{2}$ & $\mathbf{0.90}$ & $0.89$ & $0.74$ & $0.83$ & -- \\
& MAE & $\mathbf{3.83\times10^{-2}}$ & $4.00\times10^{-2}$ & -- & $5.10\times10^{-2}$ & $5.30\times10^{-2}$ \\
& RMSE & $\mathbf{5.82\times10^{-2}}$ & $6.09\times10^{-2}$ & $9.28\times10^{-2}$ & -- & -- \\
\midrule

\multicolumn{7}{l}{\textbf{Self-diffusion properties}} \\
\midrule

\multirow{3}{*}{$D_{\mathrm{N_2}}$, 1 bar}
& $R^{2}$ & $\mathbf{0.81}$ & $0.80$ & $0.76$ & $0.77$ & -- \\
& MAE & $\mathbf{3.27\times10^{-5}}$ & $3.33\times10^{-5}$ & -- & $4.52\times10^{-5}$ & $4.53\times10^{-5}$ \\
& RMSE & $\mathbf{4.46\times10^{-5}}$ & $4.53\times10^{-5}$ & $5.00\times10^{-5}$ & -- & -- \\
\addlinespace[2pt]

\multirow{3}{*}{$D_{\mathrm{N_2}}$, infinite dilution}
& $R^{2}$ & $\mathbf{0.81}$ & $0.81$ & $0.76$ & -- & -- \\
& MAE & $\mathbf{3.61\times10^{-5}}$ & $3.69\times10^{-5}$ & -- & -- & -- \\
& RMSE & $\mathbf{4.90\times10^{-5}}$ & $4.98\times10^{-5}$ & $5.50\times10^{-5}$ & -- & -- \\
\addlinespace[2pt]

\multirow{3}{*}{$D_{\mathrm{O_2}}$, 1 bar}
& $R^{2}$ & $\mathbf{0.81}$ & $0.81$ & $0.78$ & $0.78$ & -- \\
& MAE & $\mathbf{3.20\times10^{-5}}$ & $3.24\times10^{-5}$ & -- & $4.04\times10^{-5}$ & $3.99\times10^{-5}$ \\
& RMSE & $\mathbf{4.55\times10^{-5}}$ & $4.59\times10^{-5}$ & $4.98\times10^{-5}$ & -- & -- \\
\addlinespace[2pt]

\multirow{3}{*}{$D_{\mathrm{O_2}}$, infinite dilution}
& $R^{2}$ & $\mathbf{0.82}$ & $0.81$ & $0.74$ & -- & -- \\
& MAE & $\mathbf{3.12\times10^{-5}}$ & $3.17\times10^{-5}$ & -- & -- & -- \\
& RMSE & $\mathbf{4.23\times10^{-5}}$ & $4.30\times10^{-5}$ & $4.95\times10^{-5}$ & -- & -- \\
\bottomrule
\end{tabular*}

\vspace{4pt}
\parbox{0.96\linewidth}{\footnotesize
Values for the proposed representations are means over 100 test evaluations; the corresponding standard deviations are reported in the Supporting Information. Henry constants are reported in mol\,kg$^{-1}$\,Pa$^{-1}$, uptake in mol\,kg$^{-1}$, and self-diffusion in cm$^{2}$\,s$^{-1}$. A dash indicates that the corresponding metric was not reported in the benchmark study. Bold type marks the best reported mean value for each property and metric. The external results are contextual benchmarks because their sample counts and data partitions differ from those used here.
}
\end{table}

\begin{figure}[!htbp]
\centering
\includegraphics[width=\linewidth]{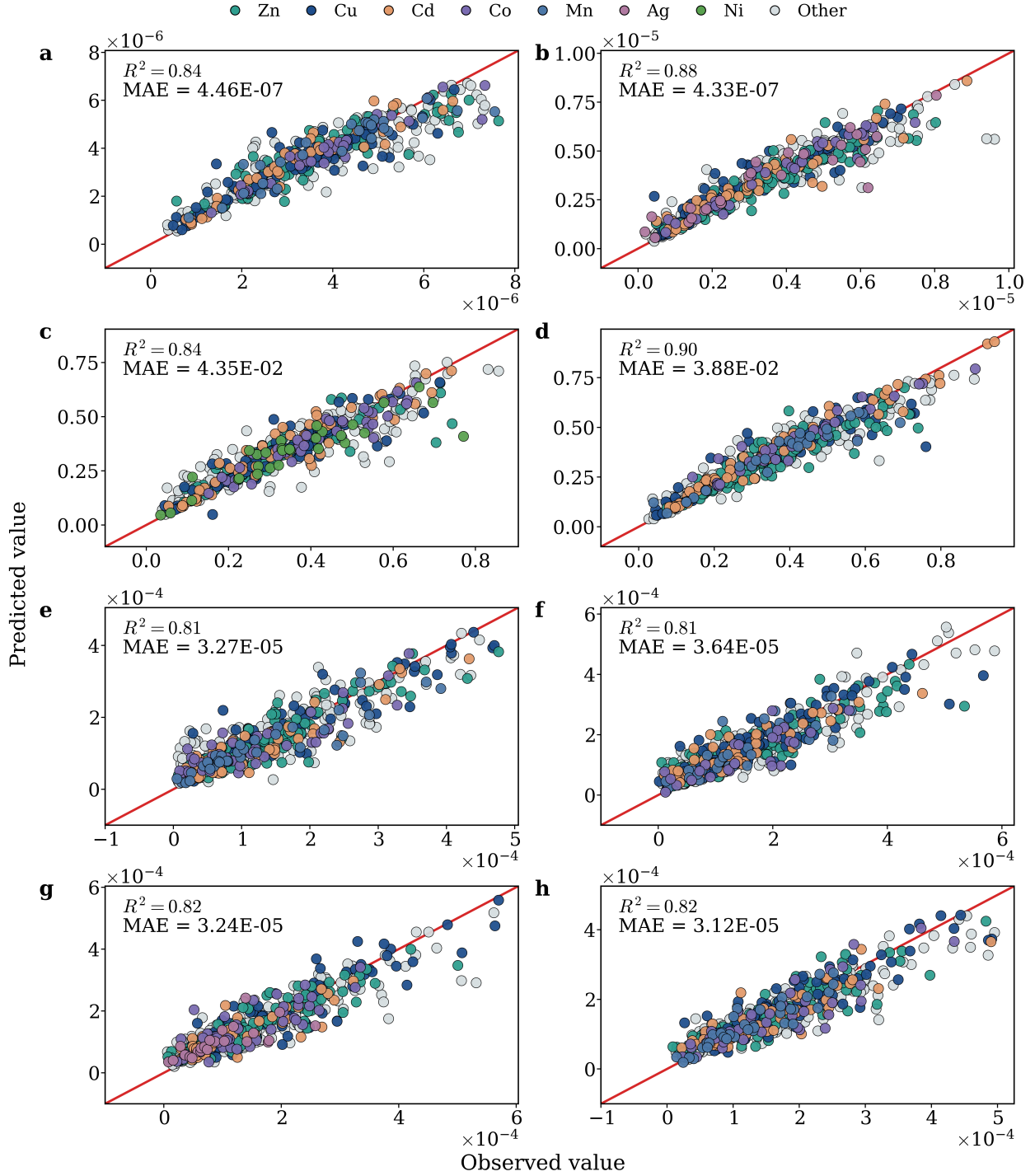}
\caption{Observed versus predicted adsorption and self-diffusion properties obtained using ERPH combined with metal coordination core persistent Mayer homology at $N=3$. Panels show \textbf{a.} N$_2$ Henry constant, \textbf{b.} O$_2$ Henry constant, \textbf{c.} N$_2$ uptake, \textbf{d.} O$_2$ uptake, \textbf{e.} N$_2$ self-diffusion at 1 bar, \textbf{f.} N$_2$ self-diffusion at infinite dilution, \textbf{g.} O$_2$ self-diffusion at 1 bar, and \textbf{h.} O$_2$ self-diffusion at infinite dilution. Each point represents one MOF and is colored by its dominant metal; the remaining metals are grouped as ``Other''. The $R^{2}$ and MAE values correspond to the displayed representative test evaluations, while Table~\ref{tab:property_benchmark} reports the mean results over 100 test evaluations. The red line represents perfect agreement between observed and predicted values.}
\label{fig:observed_predicted_n3}
\end{figure}

Figure~\ref{fig:observed_predicted_n3}a--d shows strong agreement between observed and predicted adsorption properties across MOFs with different dominant metals. Panels~\textbf{a} and \textbf{b} show that the N$_2$ and O$_2$ Henry constants are well reproduced, with stronger agreement for O$_2$. Panels~\textbf{c} and \textbf{d} show similarly strong prediction of gas uptake, with O$_2$ uptake giving the highest accuracy. Zn, Cu, Cd, Co, Mn, Ni, and Ag occur among the most frequent dominant metals, but their points overlap throughout the property ranges rather than forming separate groups. Dominant metal identity therefore does not by itself separate the adsorption ranges in these data. This observation is consistent with the combined representation retaining both coordination-core geometry and the organization of the periodically expanded structure.

Table~\ref{tab:property_benchmark} also compares the proposed representations with the available benchmark results for the four self-diffusion tasks. ERPH combined with PMH at $N=3$ gives mean $R^{2}$ values between $0.81$ and $0.82$, compared with $0.74$ to $0.78$ for the descriptor-based benchmark. The proposed representation also gives lower available MAE and RMSE values. These improvements are smaller than those observed for adsorption and should be interpreted in light of the different sample counts and partitions used in the external benchmark studies.

Figure~\ref{fig:observed_predicted_n3}e--h shows that the combined representation also captures the major structural variation associated with molecular transport in MOFs. The four self-diffusion panels give $R^{2}$ values of approximately $0.81$ to $0.82$ for N$_2$ and O$_2$ under both conditions. The major dominant-metal groups overlap across the prediction ranges, so metal identity alone does not separate the self-diffusion values. Larger deviations occur at the highest diffusion values, indicating that the fastest-diffusing structures are more difficult to predict. Overall, the combined representation captures structural variation relevant to both adsorption and molecular transport across chemically diverse MOFs.

\subsection*{Effect of Mayer order on prediction}
\label{subsec:mayer_order_results}

Table~\ref{tab:internal_ablation} shows the direct contribution of the coordination core features. Relative to ERPH, ERPH+PMHN3 increased mean test $R^2$ by $0.0381$--$0.0443$ for adsorption and $0.0134$--$0.0253$ for self-diffusion. ERPH+PMHN5 improved all eight tasks but did not exceed ERPH+PMHN3, and neither concatenation nor consensus provided a further improvement.

\begin{table}[!t]
\centering

\caption{Comparison of the evaluated topological models. Entries are mean test $R^2$ across 100 evaluations using identical data partitions and model settings.}
\label{tab:internal_ablation}

\small
\setlength{\tabcolsep}{4.5pt}
\renewcommand{\arraystretch}{1.08}

\begin{tabular}{@{}lccccc@{}}
\toprule
\textbf{Property}
& \textbf{ERPH}
& \textbf{ERPH+PMHN3}
& \textbf{ERPH+PMHN5}
& \textbf{Concatenation}
& \textbf{Consensus} \\
\midrule

Henry constant, N$_2$
& 0.80
& \textcolor{Maroon}{\textbf{0.84}}
& 0.83
& 0.84
& 0.84 \\

Henry constant, O$_2$
& 0.84
& \textcolor{Maroon}{\textbf{0.88}}
& 0.86
& 0.87
& 0.87 \\

N$_2$ uptake
& 0.80
& \textcolor{Maroon}{\textbf{0.83}}
& 0.82
& 0.83
& 0.83 \\

O$_2$ uptake
& 0.86
& \textcolor{Maroon}{\textbf{0.90}}
& 0.89
& 0.89
& 0.89 \\

$D_{\mathrm{N}_2}$, 1 bar
& 0.80
& \textcolor{Maroon}{\textbf{0.81}}
& 0.80
& 0.80
& 0.81 \\

$D_{\mathrm{N}_2}$, infinite dilution
& 0.80
& \textcolor{Maroon}{\textbf{0.81}}
& 0.81
& 0.81
& 0.81 \\

$D_{\mathrm{O}_2}$, 1 bar
& 0.80
& \textcolor{Maroon}{\textbf{0.81}}
& 0.81
& 0.81
& 0.81 \\

$D_{\mathrm{O}_2}$, infinite dilution
& 0.80
& \textcolor{Maroon}{\textbf{0.82}}
& 0.81
& 0.82
& 0.82 \\

\bottomrule
\end{tabular}

\end{table}
Paired analysis confirmed the ERPH+PMHN3 improvement across all ten splits for every property (Holm-adjusted $p\leq0.00983$). For ERPH+PMHN5, the gains were significant for adsorption and O$_2$ self-diffusion, but not for N$_2$ self-diffusion ($p=0.137$). The all-heavy-atom channel was the strongest individual PMHN3 channel, although the optimal channel combination varied by property. Complete statistical and channel-ablation results are provided in Tables~S5 and~S6, with their structural interpretation discussed in the subsection \textit{Chemical Interpretation and Connection to POM-Based MOFs}.
%~\ref{subsec:chemical_interpretation}.
\FloatBarrier

\subsection*{Topological invariants beyond those of classical persistent homology}
\label{subsec:algebraic_resolution_discussion}

\begin{figure}[H]
\centering
\includegraphics[width=\linewidth]{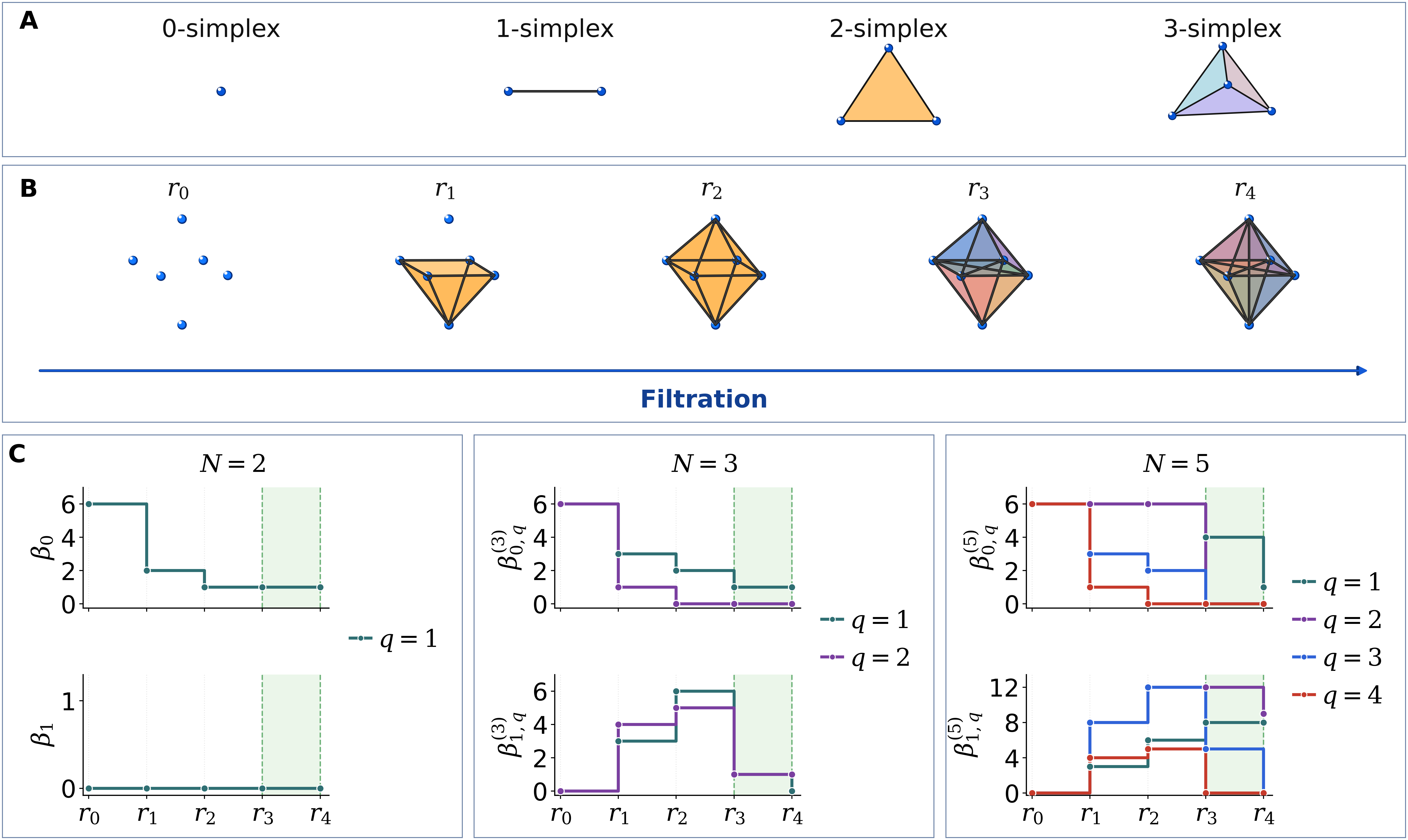}
\caption{Comparison of classical persistent homology and persistent Mayer homology on a six-point Vietoris--Rips filtration. \textbf{A.} Basic simplices from dimensions zero through three, showing the vertices, edges, triangles, and tetrahedra from which the filtered complexes are constructed. \textbf{B.} Vietoris--Rips complexes at the critical radii $r_0=0$, $r_1=\sqrt{2}$, $r_2=\sqrt{2.69}$, $r_3=2$, and $r_4=2.3$. Increasing the filtration radius progressively introduces edges and higher-dimensional simplices and changes their incidence relations. \textbf{C.} Ordinary Betti curves for $N=2$ and Mayer Betti curves for $N=3$ and $N=5$. The green region highlights the interval from $r_3$ to $r_4$. Over this interval, the ordinary $\beta_0$ and $\beta_1$ values shown here remain unchanged, whereas several Mayer Betti values change. The example therefore illustrates how different simplicial arrangements can have the same ordinary connectivity and loop summary but remain distinguishable through generalized Mayer boundary relations. The point configuration follows Example~2 of Shen et al.~\cite{shen2024}; the complexes and Betti values were calculated independently in this study. This Vietoris--Rips example illustrates the additional topological invariants provided by PMH and is not part of the MOF feature-generation procedure.}
\label{fig:vr_ph_pmh}
\end{figure}

PMH provides families of topological invariants that are not available from classical persistent homology. To isolate the information introduced by PMH from the ERPH representation and filtration used for the MOF descriptors, we consider the six-point Vietoris--Rips filtration in Fig.~\ref{fig:vr_ph_pmh}. This example compares the topological information returned by classical persistent homology and persistent Mayer homology. All MOF features used in model training were calculated from $\alpha$-complex filtrations.

Figure~\ref{fig:vr_ph_pmh} illustrates the simplices, filtration, and topological invariants obtained from classical persistent homology at $N=2$ and PMH at $N=3$ and $N=5$. Panel~\textbf{A} introduces the simplices that form the filtered complexes, while panel~\textbf{B} shows how these simplices accumulate as the filtration radius increases. By $r_3$ and $r_4$, both complexes are connected and contain no independent unfilled one-dimensional loop. Classical persistent homology therefore gives $\beta_0=1$ and $\beta_1=0$ at both stages. The complexes are nevertheless not simplicially identical because additional higher-dimensional simplices change the incidence relations among their edges, triangles, and tetrahedra. Panel~\textbf{C} shows that Mayer homology remains sensitive to this reorganization. For $N=3$, $\beta_{1,1}^{(3)}$ decreases from $1$ to $0$ between $r_3$ and $r_4$. For $N=5$, $\beta_{0,1}^{(5)}$ decreases from $4$ to $1$, while $\beta_{1,2}^{(5)}$ and $\beta_{1,3}^{(5)}$ decrease from $12$ to $9$ and from $5$ to $0$, respectively. These invariants should not be interpreted as additional ordinary holes. They measure chain-equivalence classes obtained from different powers of the Mayer boundary operator. This example shows that PMH can distinguish internal simplicial organizations that ordinary $H_0$ and $H_1$ summarize identically. These additional invariants motivate the use of PMH for metal coordination cores, where similar chemical constituents can form different internal atomic arrangements.

\subsection*{Chemical interpretation and connection to POM based MOFs}
\label{subsec:chemical_interpretation}

The predictive gains give the algebraic distinction in Fig.~\ref{fig:vr_ph_pmh} a structural meaning. ERPH describes chemically resolved organization across the periodic structure, whereas PMH resolves atomic arrangements around individual metal centers. The potential-donor, carbon, and all-heavy-atom channels describe coordinating atoms, nearby linker structure, and the complete nonhydrogen environment, respectively. PMH can therefore distinguish coordination cores with similar composition but different internal atomic arrangements. The ablation results support this complementary role. ERPH+PMHN3 improved $R^2$ on all ten splits for every property, and the all-heavy-atom channel was the strongest individual channel. ERPH+PMHN5 also improved the mean results, although its paired gains were less consistent. Neither feature concatenation nor prediction consensus surpassed ERPH+PMHN3. These findings associate the predictive improvement with the complete nonhydrogen geometry surrounding the metal centers, although the nested channels prevent their effects from being interpreted as additive. Complete paired and channel-ablation results are reported in Tables~S5 and~S6.

This distinction is particularly relevant to POM based MOFs, whose inorganic components contain multiple metal centers connected through structured oxygen environments \cite{du2014recent,zhang2021pommof,chen2023pomframeworks,ebrahimi2024pommofs}. PMH can describe the internal arrangement of these multimetal oxide cores, while ERPH retains the organization of the surrounding periodic pore structure. Their combination is therefore suited to cases in which similar cores occur within different pore environments or internal core distortions occur without major changes in pore topology. The present benchmark is not specific to POM based MOFs and does not encode oxidation state, protonation, charge, or terminal and bridging oxygen environments \cite{long2010polyoxometalates,li2019recent}. A curated POM based MOF dataset will be required to evaluate this application and determine whether the additional $N=5$ invariants become more informative for structurally complex inorganic cores.

\section*{Conclusion}
\label{sec:conclusion}
We combined persistent Mayer homology of metal-centered coordination cores with ERPH of the periodically expanded MOF structure. Across eight N$_2$ and O$_2$ adsorption and self-diffusion tasks, both Mayer orders improved mean test performance relative to ERPH alone. ERPH+PMHN3 achieved the highest mean $R^2$ and the lowest mean MAE and RMSE for every property. It also improved $R^2$ over ERPH on all ten paired data splits for every task, with statistically significant gains after Holm correction. The channel ablation clarifies the structural origin of these improvements. The all-heavy-atom channel was the strongest individual channel for all eight properties, indicating that the broader nonhydrogen geometry surrounding each metal center contains important predictive information. Carbon and potential-donor channels provide more chemically focused descriptions, but their contributions are nested within the all-heavy-atom channel and are not additive. The optimal channel combination varied among properties, showing that adsorption and diffusion depend on different aspects of the coordination-core geometry. Persistent Mayer homology therefore complements ERPH by resolving internal simplicial relations that are not retained by ordinary Betti numbers alone. This representation is particularly relevant to POM-based MOFs, where multinuclear metal oxide cores are embedded within extended porous structures. Each metal atom in a POM unit can serve as an anchor, allowing PMH to describe the internal metal and oxygen arrangement while ERPH captures the surrounding periodic pore organization. This combined representation could provide a structural basis for relating POM-core geometry, metal substitution, core distortion, and framework attachment to material properties. The present benchmark is not POM-specific and does not encode oxidation state, charge, bond order, spin state, or adsorption energetics. Future studies using curated POM-based MOF datasets and chemically resolved terminal and bridging oxygen environments will be needed to establish this broader applicability.

\section*{Methods}
\label{sec:methods}

\subsection*{Datasets}
\label{subsec:datasets}

Crystal structures were obtained from the all solvent removed CoRE MOF 2019 database \cite{chung2019core}, while the eight oxygen and nitrogen properties were taken from Orhan et al.~\cite{orhan2022}. These properties comprise the Henry constants, finite pressure uptakes, and self diffusion coefficients at 1 bar and infinite dilution for N$_2$ and O$_2$. We used the released filtered targets without additional response based filtering and matched them to the crystal structures using normalized MOF reference codes. Depending on property availability, the datasets contain 4,744 to 5,241 MOFs, as detailed in Supporting Information Table~S1. The descriptor model \cite{orhan2022}, MOFTransformer \cite{kang2023}, and PMTransformer \cite{park2023} were used as contextual benchmarks. Each dataset was divided into $80\%$ training, $10\%$ validation, and $10\%$ test subsets, with identical partitions used for all representations within each evaluation.

\subsection*{Simplicial filtrations and persistent Mayer homology}

A filtration is an ordered sequence of simplicial complexes
\begin{equation}
K_{t_0}\subseteq K_{t_1}\subseteq\cdots\subseteq K_{t_L},
\qquad
t_0<t_1<\cdots<t_L,
\label{eq:simplicial_filtration}
\end{equation}
in which vertices, edges, triangles, and higher dimensional simplices enter as the scale $t$ increases. Vietoris--Rips, \v{C}ech, and $\alpha$-complex filtrations differ in how these simplices are constructed from a point cloud \cite{hausmann1995vietoris,ghrist2014elementary,edelsbrunner1995smooth}. The theoretical example in Fig.~\ref{fig:vr_ph_pmh} uses a Vietoris--Rips filtration, whereas all MOF features are computed from $\alpha$-complex filtrations.

Persistent Mayer homology generalizes persistent homology by replacing the ordinary condition $d^2=0$ with $d^N=0$ \cite{shen2024,feng2025mayer}. An $N$-chain complex over $\mathbb{C}$ consists of
\begin{equation}
C_*=\bigoplus_{n\geq0}C_n,
\qquad
d_n:C_n\longrightarrow C_{n-1},
\qquad
d^N=0,
\label{eq:n_differential}
\end{equation}
where $C_n$ is generated by the oriented $n$-simplices and $d$ is the generalized boundary operator. For the primitive $N$th root of unity
\begin{equation}
\omega_N=\exp\left(\frac{2\pi\mathrm{i}}{N}\right),
\end{equation}
the boundary of an oriented simplex is
\begin{equation}
d_n[v_0,\ldots,v_n]
=
\sum_{j=0}^{n}
\omega_N^{\,j}
[v_0,\ldots,\widehat{v_j},\ldots,v_n],
\label{eq:mayer_boundary}
\end{equation}
where $\widehat{v_j}$ denotes omission of $v_j$. When $N=2$, $\omega_2=-1$, and this expression reduces to the ordinary alternating boundary operator.

For $q\in\{1,\ldots,N-1\}$, the relation
$d^q d^{N-q}=d^N=0$ gives the Mayer cycle and boundary spaces
\begin{align}
Z_n^{N,q}(C_*,d)
&=
\ker\left(d^q:C_n\longrightarrow C_{n-q}\right),
\label{eq:mayer_cycles}
\\
B_n^{N,q}(C_*,d)
&=
\operatorname{im}\left(
d^{N-q}:C_{n+N-q}\longrightarrow C_n
\right).
\label{eq:mayer_boundaries}
\end{align}
Since $B_n^{N,q}\subseteq Z_n^{N,q}$, the Mayer homology group and Mayer Betti number are
\begin{align}
H_n^{N,q}(C_*,d)
&:=
\frac{Z_n^{N,q}(C_*,d)}
{B_n^{N,q}(C_*,d)},
\label{eq:mayer_homology}
\\
\beta_n^{N,q}(C_*,d)
&:=
\dim H_n^{N,q}(C_*,d).
\label{eq:mayer_betti}
\end{align}
Ordinary homology is recovered when $N=2$ and $q=1$. For $N>2$, the different values of $q$ resolve additional relations among the vertices, edges, triangles, and higher dimensional simplices of the same complex.

Let $f:K\rightarrow\mathbb{R}$ satisfy
\begin{equation}
\tau\subseteq\sigma
\quad\Longrightarrow\quad
f(\tau)\leq f(\sigma),
\end{equation}
and define
\begin{equation}
K_r=\{\sigma\in K:f(\sigma)\leq r\}.
\end{equation}
For $r\leq s$, the inclusion $K_r\hookrightarrow K_s$ induces a map on Mayer homology, and the persistent Mayer group is
\begin{equation}
H_n^{N,q}(r,s)
:=
\operatorname{im}\left[
H_n^{N,q}\bigl(C_*(K_r),d\bigr)
\longrightarrow
H_n^{N,q}\bigl(C_*(K_s),d\bigr)
\right].
\label{eq:persistent_mayer_homology}
\end{equation}
The features used here are obtained from the corresponding Mayer Betti curves
\begin{equation}
\beta_n^{N,q}(r)
:=
\dim H_n^{N,q}\bigl(C_*(K_r),d\bigr).
\label{eq:mayer_betti_curve}
\end{equation}
These curves measure scale dependent changes in the $q$ indexed boundary relations. They are not additional counts of ordinary connected components, loops, or cavities.

\subsection*{Construction of ERPH and coordination core features}
\label{subsec:feature_construction}

All MOF representations are constructed from atomic coordinates using $\alpha$-complex filtrations. For an atomic point cloud
\begin{equation}
P=\{\mathbf{x}_1,\ldots,\mathbf{x}_m\}\subset\mathbb{R}^3,
\end{equation}
let $\operatorname{Del}(P)$ be its Delaunay triangulation and let $\rho(\sigma)$ be the radius of the smallest empty circumsphere of $\sigma$. The $\alpha$-complex is
\begin{equation}
K_{\alpha}(P)
=
\left\{
\sigma\in\operatorname{Del}(P):
\rho(\sigma)\leq\alpha
\right\}.
\label{eq:alpha_complex}
\end{equation}
Thus, $K_{\alpha_i}(P)\subseteq K_{\alpha_j}(P)$ whenever $\alpha_i\leq\alpha_j$. The resulting simplices describe spatial relations among atoms and do not require predefined chemical bonds. Each periodic MOF was expanded along lattice direction $i$ using
\begin{equation}
n_i=
\max\left\{
1,
\operatorname{round}
\left(
\frac{64}{\lVert\mathbf{a}_i\rVert}
\right)
\right\},
\end{equation}
where $\mathbf{a}_i$ is the corresponding lattice vector. This produces lattice vector lengths approaching $64~\text{\AA}$ without rescaling the unit cell or imposing cubic symmetry. The expansion provides a common spatial domain while preserving the lattice geometry and periodic atomic connectivity of each MOF.

For element resolved persistent homology, the atoms are partitioned into
\begin{equation}
\mathcal{C}
=
\left\{
C_m^{(1)},C_m^{(2)},C_n,C_o,C_p,C_q,C_r,C_s,C_{\Omega}
\right\}.
\end{equation}
Here, $C_m^{(1)}$ contains main group metals, $C_m^{(2)}$ contains transition and inner transition metals, $C_n$ contains metalloids, $C_o$ contains halogens, $C_p$ contains H, $C_q$ contains C, $C_r$ contains N and P, $C_s$ contains O, S, and Se, and $C_{\Omega}$ contains the complete structure. These groups separate the metal nodes, linker atoms, potential coordination atoms, and complete atomic arrangement.

For each $C\in\mathcal{C}$, let $P_C$ be the corresponding atomic point cloud and define
\begin{equation}
K_{\alpha}^{C}=K_{\alpha}(P_C).
\end{equation}
The element resolved Betti curves are
\begin{equation}
\beta_k^{C}(\alpha)
=
\dim H_k\left(K_{\alpha}^{C}\right),
\qquad
k=0,1,2.
\label{eq:erph_betti_curve}
\end{equation}
They are evaluated on
\begin{equation}
\mathcal{A}_{\mathrm{ERPH}}
=
\left\{
\alpha_{\ell}=0.1\ell:
\ell=0,\ldots,249
\right\}\ \text{\AA},
\label{eq:erph_alpha_grid}
\end{equation}
giving the feature set
\begin{equation}
\mathcal{F}_{\mathrm{ERPH}}
=
\left\{
\beta_k^{C}(\alpha_{\ell}):
C\in\mathcal{C},\
k\in\{0,1,2\},\
\alpha_{\ell}\in\mathcal{A}_{\mathrm{ERPH}}
\right\}.
\label{eq:erph_features}
\end{equation}

Persistent Mayer homology is applied to the coordination environment surrounding each metal atom. Let $a$ be a metal anchor at $\mathbf{x}_a$, with its element belonging to $C_m^{(1)}\cup C_m^{(2)}$. Every metal site is treated as an anchor, and its element label distinguishes different metal species within the same MOF. For channel $c$, the periodic metal centered point cloud is
\begin{equation}
P_{a,c}
=
\{\mathbf{0}\}
\cup
\left\{
\mathbf{x}_j+\boldsymbol{\lambda}-\mathbf{x}_a:
\left\|
\mathbf{x}_j+\boldsymbol{\lambda}-\mathbf{x}_a
\right\|
\leq10~\text{\AA},
\quad
e_j\in E_c,
\quad
(j,\boldsymbol{\lambda})\neq(a,\mathbf{0})
\right\},
\label{eq:metal_core_point_cloud}
\end{equation}
where $\boldsymbol{\lambda}$ is a lattice translation. The anchor occupies the origin, and all periodic images satisfying the cutoff and channel condition are retained.

The three chemical channels are
\begin{align}
E_{\mathrm{AH}}
&=
\{e:e\neq\mathrm{H}\},
\label{eq:heavy_atom_set}
\\
E_{\mathrm{C}}
&=
\{\mathrm{C}\},
\label{eq:carbon_set}
\\
E_{\mathrm{D}}
&=
\{\mathrm{N},\mathrm{O},\mathrm{S},\mathrm{Se},\mathrm{P},
\mathrm{F},\mathrm{Cl},\mathrm{Br},\mathrm{I}\}.
\label{eq:donor_set}
\end{align}
The all heavy atom channel describes the complete nonhydrogen arrangement around the metal, including neighboring metal atoms. The carbon channel emphasizes the surrounding linker scaffold, while the potential donor channel emphasizes atoms that may participate in metal ligand coordination. Chemistry is therefore introduced through the anchor and partner element labels, while the filtration describes their spatial organization.

To identify repeated coordination configurations, pairwise distances are quantized using
\begin{equation}
\widehat d_{ij}
=
\operatorname{round}
\left(
\frac{d_{ij}}{\varepsilon}
\right),
\qquad
\varepsilon=10^{-6}~\text{\AA}.
\label{eq:quantized_pairwise_distance}
\end{equation}
Coordinates are deduplicated at $10^{-8}~\text{\AA}$. A canonical code is formed from the anchor or partner roles, element labels, and upper triangular quantized distance matrix under all role and element preserving permutations. Point clouds with identical canonical codes define the same coordination core class. This removes differences caused only by atom ordering or periodic image selection while preserving chemical identity and complete internal geometry.

Let $\mathcal{U}_c$ denote the distinct coordination core classes in channel $c$. Each class contributes once, irrespective of how many equivalent metal sites occur in the crystal. Invariance under Euclidean transformations, label preserving permutations, and periodic representative choices is described in Supporting Information Section under Periodic construction and invariance of the coordination cores.
%~\ref{sec:si_periodic_core_consistency}.

For representative $u\in\mathcal{U}_c$, define
\begin{equation}
K_{\alpha}^{u,c}=K_{\alpha}(P_{u,c}),
\end{equation}
and compute
\begin{equation}
\beta_{k,q}^{N,u,c}(\alpha)
=
\dim
H_k^{N,q}
\left(
C_*\left(K_{\alpha}^{u,c}\right),d
\right).
\label{eq:coordination_core_mayer_betti}
\end{equation}
These curves resolve how the atoms within a metal coordination core are connected through the different powers of the Mayer boundary operator.

The curves are sampled on
\begin{equation}
\mathcal{A}_{\mathrm{core}}
=
\left\{
\alpha_{\ell}=0.2\ell:
\ell=0,\ldots,39
\right\}\ \text{\AA}.
\label{eq:coordination_core_alpha_grid}
\end{equation}
For each chemical channel, the distinct coordination core classes are summarized by
\begin{align}
\beta_{k,q,\max}^{N,c}(\alpha_{\ell})
&=
\max_{u\in\mathcal{U}_c}
\beta_{k,q}^{N,u,c}(\alpha_{\ell}),
\label{eq:coordination_core_max}
\\
\beta_{k,q,\mathrm{mean}}^{N,c}(\alpha_{\ell})
&=
\frac{1}{|\mathcal{U}_c|}
\sum_{u\in\mathcal{U}_c}
\beta_{k,q}^{N,u,c}(\alpha_{\ell}).
\label{eq:coordination_core_mean}
\end{align}
Both statistics are unweighted over distinct classes. We use $N=3$ with $q=1,2$ and $N=5$ with $q=1,2,3,4$, in dimensions $k=0,1$.

\noindent\textbf{Predictive modeling and evaluation.}
Gradient boosting decision tree regression was used for all eight properties \cite{friedman2001greedy,pedregosa2011scikit}. Ten data split seeds and ten model seeds produced 100 evaluations per representation, with identical partitions and seed pairs used throughout. Test performance was measured using $R^2$, MAE, and RMSE. The evaluated models included ERPH, ERPH+PMHN3, ERPH+PMHN5, feature concatenation, and prediction consensus. Model hyperparameters, metric definitions, and statistical procedures are given in the Supporting Information.

\section*{Data and Code Availability}

The MOF crystal structures were obtained from the all-solvent-removed collection of the CoRE MOF 2019 database \cite{chung2019core}. The oxygen and nitrogen adsorption and self-diffusion property data were obtained from the datasets reported by Orhan et al.~\cite{orhan2022}. The code and computational resources supporting this work are maintained at
\url{https://github.com/CSCA-MOFs/MOF-PMH-and-PH}. 

\section*{Supporting Information}
Dataset composition and benchmark sample counts; prediction metrics and statistical evaluation; feature-construction settings; periodic coordination-core construction and invariance proofs; numerical verification of the persistent Mayer homology calculations; complete representation ablation; paired split analysis; and coordination-channel ablation are included in the Supporting Information.

\section*{Acknowledgments}

This work was supported in part by the University of Georgia and Georgia Research Alliance to G.-W. W. C.-L. C. and G.-W. W. also acknowledge financial support from DOE, Office of Science, BES under an award FWP 84274 at Pacific Northwest National Laboratory (PNNL) for the development of machine learning models. PNNL is a multiprogram national laboratory operated for the Department of Energy by Battelle under Contract DE-AC05-76RL01830.

\bibliographystyle{unsrt}
\bibliography{reference}

\end{document}